\documentclass{ar-1col-S2O}
\usepackage[numbers]{natbib}
\usepackage{url}
\usepackage{bm}
\usepackage{braket}
\usepackage{amsmath,amssymb}

\usepackage{hyperref} 
\usepackage{xspace}
\usepackage[normalem]{ulem}

\newcommand\rout{\bgroup\markoverwith
      {\textcolor{red}{\rule[.3ex]{2pt}{1.5pt}}}\ULon}

\begin{document}

\markboth{Jonathan Engel}{Schiff Moments}

\title{Nuclear Schiff Moments and CP Violation}

\author{Jonathan Engel$^1$ 
\affil{$^1$Department of Physics and Astronomy, University of North Carolina,
Chapel Hill, NC, 27516, USA, ; email: engelj@physics.unc.edu ; 
ORCID: 0000-0002-2748-6640}
}

\begin{abstract}
This paper reviews the calculation of nuclear Schiff moments, which one must
know in order to interpret experiments that search for time-reversal-violating
electric dipole moments in certain atoms and molecules.  After briefly reviewing
the connection between dipole moments and CP violation in and beyond the
Standard Model of particle physics, Schiff's theorem, which concerns the
screening of nuclear electric dipole moments by electrons, Schiff moments, and
experiments to measure dipole moments in atoms and molecules, the paper examines
attempts to compute Schiff moments in nuclei such as $^{199}$Hg and
octupole-deformed isotopes such as $^{225}$Ra, which are particularly useful in
experiments.  It then turns to \textit{ab initio} nuclear-structure theory,
describing ways in which both the In-Medium Similarity Renormalization Group and
coupled-cluster theory can be used to compute important Schiff moments more
accurately than the less controlled methods that have been applied so far.
\end{abstract}

\maketitle

\tableofcontents


\section{INTRODUCTION}
\label{sec:intro}

It is a strange time for the field of fundamental physics.  There is no shortage
either of unexplained phenomena or of theoretical puzzles.  We still do not know
what dark matter is made of, we understand little about dark energy, and we
cannot explain why the universe contains so many more baryons than anti-baryons,
an imbalance that is difficult to chalk up to an initial condition if inflation
takes place \cite{Coppi04}.  Sakharov famously pointed out \cite{Sakharov91}
that a phase transition in concert with a source of CP violation could explain
the asymmetry, but by all accounts the one known source of CP violation, a phase
in the Cabibbo-Kobayashi-Maskawa (CKM) matrix that mixes quark mass and flavor
eigenstates \cite{Kobayashi73}, is too weak.  For this reason, and because
global symmetries such as CP are often not regarded as natural, many people
suspect that nature violates CP invariance more strongly than does the CKM
phase.  This additional CP-violating physics has not, of course, been seen in
high-energy experiments.  As we shall see shortly, however, static electric
dipole moments (EDMs) in atoms and molecules without a lot of degeneracy violate
time-reversal (T) symmetry, which implies CP violation \cite{Luders54}, and
experiments to measure EDMs are both exquisitely sensitive and becoming even
more sensitive at a rapid pace.  Such experiments may thus provide our best
chance to discover a new source of CP violation.

One mystery is that there is an additional potential source of CP violation in
the standard model that isn't doing doing the job it might: the term in the QCD
Lagrangian with the form
\begin{equation}
\label{eq:theta-term}
\mathcal{L}_{\bar{\theta}} = - \frac{g^2}{16\pi^2} \bar{\theta} \, {\rm Tr} \left(
G^{\mu \nu} \tilde{G}_{\mu \nu} \right) \,,
\end{equation}
where $\bar{\theta}$ includes the effects of CP-violating angles in the quark
mass matrix that we assume have been rotated away.  The QCD CP violation should
cause the neutron (as well as atoms and molecules) to have an EDM, but
experiments \cite{Abel20} show that its EDM is less than about $1.8 \times
10^{-26} e\ {\rm cm}$.  That result implies that $\bar{\theta} \lesssim
10^{-10}$, a value so small that it seems to require an explanation.  Although
many have been offered \cite{Nelson84,Barr84} --- the presence of new particles
called axions is probably the most popular \cite{Peccei77,Wilczek78,Weinberg78}
--- none have been shown to be correct.

Whatever the ultimate source of non-CKM CP violation might be, its presence
would cause atomic or molecular EDMs, the concern in this paper, that reflect
T-violating properties either of the electron or of the atomic nucleus (or of
both).  In the second case, the nucleus would induce an EDM in the system that
contains it through its interaction with the electrons.  As we shall see, the
nuclear physics responsible for this T-violating interaction can often be
summarized in what is called the nuclear ``Schiff moment,'' which is a kind of
radially weighted nuclear EDM.  Our ability to interpret an observation (or
non-observation) of an atomic or molecular EDM depends on understanding the
dependence of the nuclear Schiff moment on the underlying source of CP
violation.  Though particle theory, QCD, and effective field theory are all
required to make the connection, nuclear-structure theory is particularly
important because the uncertainty associated with its methods is particularly
large and because the field is poised to reduce that uncertainty significantly.
The delicate nuclear structure that affects Schiff moments will therefore be the
central topic of this review.

\section{EDMS AND CP VIOLATION}
\label{sec:edm}

Why do electric dipole moments violate T?  The argument is a little different
from the usual quantum-mechanical one that depends only on the commutation of
the Hamiltonian with a symmetry operator.  The electric-dipole operator in
quantum mechanics,
\begin{equation}
\label{eq:edm-op-general}
\bm{D} = \sum_i q_i \bm{r}_i \,, 
\end{equation}
where the index $i$ labels particles, $q_i$ is the charge of the $i^{\text{th}}$
particle, and $\bm{r}_i$ is its position vector, has negative parity.  As a
result, by the usual kind of argument, parity conservation implies that a static
dipole moment in a state without any degeneracy beyond that caused by rotational
symmetry must vanish.  Though the demonstration is a little more involved, the
conservation of time-reversal symmetry implies the same thing, with or without
parity conservation \cite{sachs87}.  

The argument goes as follows: The time-reversal operator $T$, because it
reverses angular momenta, takes normalized states with well defined angular
momentum $J$ and projection $M$ into normalized states with $J$ and $-M$.  Thus,
if time-reversal symmetry is conserved, one must have within a rotational
multiplet $\ket{J,M}$ of definite energy,
\begin{equation}
\label{eq:dip-T}
\begin{aligned}
\braket{J,M|D_z| J,M} & = \braket{J,M|T^{-1} T D_z T^{-1}
T|J,M} \\
& = \braket{J,-M|T D_z T^{-1}|J,-M} \\
&= \braket{J,-M|D_z|J,-M} \,,
\end{aligned}
\end{equation}
where the last equality holds because $\bm{D}$, which depends only on positions,
is even under time reversal.  The operator $R_\pi$ that rotates around the $x$
axis by $\pi$ also takes $\ket{J,M}$ to a phase times $\ket{J,-M}$, but but $\bm
D$ is odd under this operation.  Thus
\begin{equation}
\label{eq:dip-R}
\begin{aligned}
\braket{ J,M|D_z| J,M} 
& = \braket{J,M|R_\pi^{-1} R_\pi D_z R_\pi^{-1} R_\pi|J,M} \\
&= - \braket{J,-M|D_z|J,-M} \,.
\end{aligned}
\end{equation}
Equations \ref{eq:dip-T} and \ref{eq:dip-R} together imply that
$\braket{J,M|D_z|J,M} = 0$.  The argument breaks down if time-reversal symmetry
is violated because in that case the state $\ket{J,-M}$ in the second and third
lines of Eq.\ \ref{eq:dip-T} need not belong entirely to the same rotational
multiplet as the state $\ket{J,M}$, and thus need not be the same state as
$\ket{J,-M}$ in Eq.\ \ref{eq:dip-R}.

Because of the CPT theorem, a nonzero EDM for a state with no degeneracy beyond
that in $M$ implies that CP symmetry is violated.  Not only that, for years to
come, a detectable EDM will have to be caused by $\mathcal{L}_{\bar{\theta}}$ or
physics beyond the Standard Model, even with the amazing experimental
sensitivity that is already possible.  The reason is that the CKM phase causes a
change of flavor and so flavor-diagonal quantities such as EDMs require Feynman
diagrams with several loops to produce a non-zero result.  \textbf{Figure
\ref{fig:nedm}} below shows one of the leading diagrams \cite{Shabalin:1982sg}
in the expression for the neutron EDM, which the result of a full calculation
reveals \cite{Shindler21} to be about $10^{-32} e\ {\rm cm}$.  Experiments
looking for a new flavor-conserving source of CP violation will have to increase
their sensitivity by several orders of magnitude before background from the CKM
phase becomes an issue.

\begin{figure}[h]
\includegraphics[width=.75\textwidth]{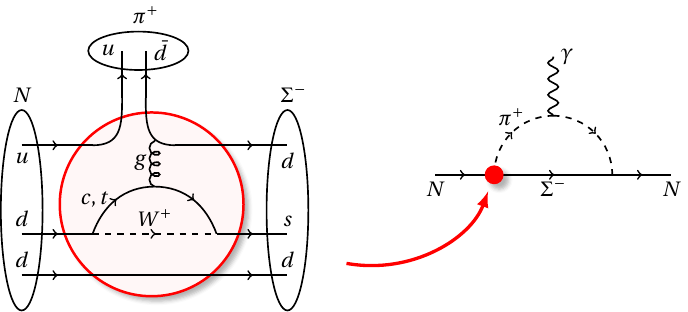}
\caption{A leading diagram in the Standard Model for the neutron EDM caused by
the CKM phase.  Lower-case letters on the left label quark flavors, $W$ labels a
charged weak boson, and the ellipses represent the collection of valence quarks
that make up the neutron ($N$), pion ($\pi$), and strange baryon ($\Sigma$). The
red circle on the right contains the corresponding larger circle on the left.
The wavy line labeled $\gamma$ represents a photon.}
\label{fig:nedm}
\end{figure}

Physics from beyond the Standard Model could produce much larger EDMs. Theories
of new physics typically generate EDMs at one or two loops, and experiments on
the neutron are already sensitive to new physics at the TeV scale for one-loop
diagrams.  Experiments on the electron EDM \cite{Andreev18} are sensitive
\cite{Alarcon22} at present to a new-physics scale of perhaps 50 TeV (for one
loop diagrams) or 2 TeV (for two loop diagrams), and, as just noted, have orders
of magnitude room to improve before background from CKM CP violation makes
discovering new physics harder.

\subsection{Beyond the CKM Matrix and Effective Couplings for Nuclear CP
Violation}
\label{sec:BSM}

We see that EDMs are sensitive to new sources of CP violation, but how do we
compute the EDMs predicted by possible new sources?  If we had to do the full
calculation for each theory of new physics we'd be in trouble, but we can use a
hierarchy of energy scales --- fundamental physics at and above the TeV scale,
the Standard Model at a hundred GeV or so, mesons and nucleons at about 1 GeV,
nuclei at the MeV scale, and atoms at the eV scale --- to divide the work so
that the missing information at each level can be summarized in a few unknown
parameters. 

Here we will be interested in atomic and molecular EDMs that are induced by
physics inside atomic nuclei.  At the nuclear level, the framework that makes a
parameterization of higher-scale physics possible is chiral effective field
theory ($\chi$EFT), which expands the most general pion and nucleon Lagrangian
that is consistent with spontaneously-broken chiral symmetry in powers of
$p/\Lambda$ and $m_\pi/\Lambda$, where $p$ is a typical momentum for a nucleon
in a nucleus, $m_\pi$ is the pion mass, and $\Lambda$ is the scale at which the
dynamics of degrees of freedom beyond pions and nucleons become important
\cite{Leutwyler94,Machleidt11,Epelbaum20}.

At leading order in $\chi$EFT, the usual strong nuclear potential contains
one-pion-exchange and contact nucleon-nucleon interactions.  The same is often
true of the leading-order P- and T-violating potential $V_{PT}$
\cite{Maekawa:2011vs,DeVries2020}, though exactly which terms are leading
depends on the underlying sources of CP violation.  With the use of the strong
pion-nucleon coupling $g \approx 13.3$ in the definition instead of $g_A m_N
/f_\pi$, which is equal to $g$ to within a few percent, the pion-exchange part
(always occurring at leading order) is
\begin{equation}
\label{eq:vpt}
\begin{aligned}
V_{PT}^\pi(\bm{r}_1-\bm{r}_2)
& =  \frac{g}{2 m_N}
         \bigg\{ \left[ \bar{g}_0 \, \vec{\tau}_1 \cdot \vec{\tau}_2
               + \bar{g}_2
               (3\tau_{1z}\tau_{2z} - \vec{\tau}_1 \cdot
               \vec{\tau}_2 ) \right] 
               (\bm{\sigma}_1 - \bm{\sigma}_2) 
                          \\
               & \qquad\qquad\qquad - \bar{g}_1 \left(
                \bm{\sigma}_1\tau_{1z}-\bm{\sigma}_2\tau_{2z} \right) 
                \bigg\} \cdot \bm{\nabla} Y(| \bm{r}_1 - \bm{r}_2|)
                \,, 
\end{aligned}
\end{equation}
where 
\begin{equation}
\label{eq:Uofr}
Y(r) = \frac{e^{-m_\pi r}}{4 \pi r} \,, 
\end{equation}
and the $\bar{g}_i$ are unknown CP-violating versions of $g$ that depend on the
underlying source of the violation.  For special sources, e.g., the
$\bar{\theta}$ term in Eq.\ \ref{eq:theta-term}, theorists have used lattice QCD
to compute the constant $\bar{g}_0$ \cite{Borsanyi15,Brantley16}, obtaining the
value $\bar{g}_0 = (15.5 \pm 2.6) \times 10^{-3} \bar{\theta}$.  The other
couplings are harder to calculate, though Ref.\ \cite{Bsaisou13} used resonance
saturation to conclude, again for the $\bar{\theta}$ source, that
$\bar{g}_1/\bar{g}_0 \approx -0.2$.

Pions are the pseudo-Goldstone bosons associated with the spontaneous breaking
of chiral symmetry, and for sources of CP violation that conserve chiral
symmetry at the quark and gluon level --- i.e.\ in Standard-Model effective
field theory --- pion-exchange potentials of the form in Eq.\ \ref{eq:vpt} are
suppressed.  For such sources, a contact interaction with two parameters
contributes at the same order as the suppressed pion exchange:  
\begin{equation}
\label{eq:vdel}
V^\delta_{PT} = \frac{1}{2} \left[ \bar{C}_1 
+ \bar{C}_2  \bm{\tau}_1 \cdot \bm{\tau}_2 \right] 
\left( \bm{\sigma}_1-\bm{\sigma}_2 \right) \cdot
\bm{\nabla}\delta^3(\bm{r_1 - \bm{r}_2}) \,.
\end{equation}
There are other contact interactions, not shown here, that never contribute at
the same order as pion exchange. In addition, according to Standard-Model EFT,
$\bar{g}_2$ is suppressed compared to $\bar{g}_0$ and $\bar{g}_1$, no matter
what the underlying source of CP violation.  For a review of EFT for P- and
T-violating interactions and operators, see Ref.\ \cite{DeVries2020}.  

The most important result of all these considerations for the interpretation of
experiments on atoms or molecules is that we can proceed to compute the effects
of nuclear CP violation on EDMs as functions of a few important $\chi$EFT
parameters, without worrying about the underlying source of CP violation.  We
will see how to do so shortly.

\section{EDMS of ATOMS AND MOLECULES}
\label{sec:atom-mol}

Atoms and molecules are easier to manipulate in a laboratory than lone hadrons
or leptons.  If CP is violated, the EDMs of those hadrons and leptons will
contribute to an EDM for the composite system.  But the charge distribution of a
composite system is also altered by the EDMs of its constituents, and one can
ask whether the dipole moment of the new distribution will act to increase or
reduce the summed EDMs of the constituents.  The answer depends on whether the
constituents are, like electrons, relativistic and spread out over the
atomic/molecular volume or, like the nuclear constituents, confined to a much
smaller volume.  In the former case, the constituent EDMs can be enhanced, but
in the latter they are dramatically screened.  We will see how next.

\subsection{Schiff's Theorm and Schiff Moments}
\label{sec:theorem}

\indent

Why are Schiff moments important?  The reason is that nuclear EDMs are largely
canceled by atomic electrons, which re-arrange themselves to create an EDM in
the opposite direction.  This result was first proved, not surprisingly, by
Schiff in Ref.\ \cite{schiff63} and is presented in many other places (see,
e.g., Refs.\ \cite{khriplovich97,Senkov08} The proof here follows the discussion
in Ref.\ \cite{engel00}, where more details are available.  One begins by
supposing that the nucleus has a dipole moment $\bm{d} = -e \bm{d}_0 $, where
$e$ is the charge of the electron (the negative of the charge of the proton) and
$d_0$ is a nuclear quantity --- a ground-sate expectation value --- that has the
dimension of length.  With this assumption, one can write the atomic Hamiltonian
in the form 
\begin{equation}
\label{eq:atom}
H_{\rm atom} = \sum_{i=1}^Z \left[ T_i + V_i + e \phi (\bm{r}_i) - 
e\bm{E}_0 \cdot \bm{r}_i \right] +e \bm{E}_0 \cdot \bm{d}_0 \ \ , 
\end{equation}
where $T_i$ is the kinetic energy of the $i^{\text{th}}$ electron ($p^{2}_i/2
m_e$ in the nonrelativistic approximation, with $m_e$ the electron mass), and
$\bm{r}_i$ and $\bm{p}_i$ are that electrons' coordinates and momentum relative
to the nuclear center of mass.  In addition, 
\begin{equation}
V_i = e^2 \sum_{j < i} \frac{1}{|\bm{r}_i - \bm{r}_j|} 
\end{equation}
is the electron-electron Coulomb interaction, and
\begin{equation}
\phi(\bm{r}_i) = -e \int \frac{d^3 x \, \rho(\bm{x})}
{|\bm{x} - \bm{r}_i|} 
\end{equation}
is the electrostatic potential due to the nuclear charge distribution $\rho
(\bm{x})$, which is dimensionless and normalized to $Z$.  One needn't worry
about the internal nuclear Hamiltonian; $\rho(\bm{x})$ is just a
nuclear-ground-state charge distribution.  Virtual nuclear excitations turn out
not to affect the conclusions reached here. 

Now, transform the Hamiltonian by using the unitary operator
\begin{equation}
U = e^{iA} \,, \quad  A = \frac{\bm{d}_0}{Z} \cdot \sum_{i = 1}^{Z} \bm{p}_i  \,.
\end{equation}
Because $d_0$ is so small, the transformed Hamiltonian $\bar{H}_{\rm atom}$, to
the level of accuracy one needs, is $\bar{H}_{\rm atom} \equiv U H U^{-1} \simeq
H_{\rm atom} + i[A, H_{\rm atom}]$. The expectation value of the commutator
$[A,H_{\rm atom}]$ must vanish in any eigenstate of $H_{\rm atom}$.  Because the
commutator of the momentum operator with of a function of $\bm{r}$ is
proportional to the gradient of that function, because the electric field is the
negative of the gradient of the potential, and because the sum of the nuclear
field vectors at the locations of all the electrons is the negative of their
combined field $\bm{E}_e$ at the center of the nucleus, the ground-state
expectation value of $[A,H_{\rm atom}]$, in the limit that the nucleus is a
point particle, gives
\begin{equation}
\label{eq:}
\braket{\bm{d}_0 \cdot \left[ \bm{E}_0 + \bm{E}_e \right]} = 0 \,.
\end{equation}
Finally, because we could have used any vector in place of $\bm{d}_0$ in the
operator $A$, this means that the external field causes the electrons to move so
that the field they produce exactly cancels the external field at the nucleus.
Thus, the point-nucleus's EDM does not affect the total energy to first order
when the external field is turned on.  The nuclear EDM is screened.

To see how the finite volume of the nucleus affects this screening, one can
examine the transformed Hamiltonian $\bar{H}_{\text{atom}}$: 
\begin{equation}
\label{hbaratom}
\begin{aligned}
\bar{H}_{\rm atom} &= \sum^{Z}_{i=1} \left(T_i + V_i - e \bm{E}_0 \cdot 
\bm{r}_i  + e \bar{\varphi}(\bm{r}_i) \right) \\
\bar{\varphi}(\bm{r}) & =  \phi (\bm{r}) + \frac{1}{Z} \, \bm{d}_0 \cdot 
\bm{\nabla} \, \phi (\bm{r}) \, . 
\end{aligned}
\end{equation}
Although the internal nuclear Hamiltonian does not appear here, the nuclear
radius $R_N$ is still a relevant parameter because it characterizes
$\rho(\bm{x})$, and it is much smaller than atomic radius $R_A$ .  Thus one can
expand $\rho(\bm{x})$ in powers of $R_N/R_A \approx 10^{-4}$ and neglect all but
the lowest few terms.  The expansion is equivalent to an expansion in gradients
of $\delta^3(\bm{x})$.  In order to reproduce the lowest multipole moments, we
write
\begin{equation}
\label{eq:density-expansion}
\rho (\bm{x}) = \left[ Z \, \delta^3 (\bm{x}) + Z \frac{\langle r^2
\rangle_{\rm ch}}{6} \bm{\nabla}^2 \, \delta^3 ({\bm{x}}) + \cdots \right] - \left[
\bm{d}_0 \cdot \bm{\nabla} \delta^3 ({\bm{x}}) + \frac{\bm{O}_0 \cdot
\bm{\nabla}}{10} \bm{\nabla}^2 \, \delta^3 ({\bm{x}}) + \cdots \right] 
+ \cdots  \,, 
\end{equation}
where $\braket{r^2}_{\rm ch}$ is the mean-square charge radius and
\begin{equation}
\bm{O}_0 = \int d^3 x\, \bm{x}\, x^2 \, \rho (\bm{x}) \,. 
\end{equation}
The vector quantity $\bm{O}_0$, which is the second moment of the dipole
distribution, bears a similar relationship to $\bm{d}_0$ as $Z \langle r^2
\rangle_{\rm ch}$ does to $Z$.  The terms in the second set of square brackets
and higher terms with odd multipoles exist only if the nuclear Hamiltonian
violates parity and time-reversal symmetry. 

Now one can use the expression for the density in Eq.\
\ref{eq:density-expansion} to evaluate the modified potential $\bar{\varphi}$ in
Eq.\ \ref{hbaratom}.  One can write the potential as $\bar{\varphi} \equiv
\bar{\varphi}_0 + \bar{\varphi}_{PT}$, a sum of a term $\bar{\varphi}_0$ that
would be present even in the absence of nuclear P and T violation and another
symmetry-violating term $\bar{\varphi}_{PT}$ that contains $\bm{d}_0$ or
$\bm{O}_0$, obtaining 
\begin{equation}
\label{eq:Del}
\begin{aligned}
\bar{\varphi}_{PT} (\bm{r}) &= e \int \! d^3x \, \frac{\left[
\bm{d}_0 \cdot \bm{\nabla} \delta^3 ({\bm{x}}) + \frac{1}{10} \bm{O}_0 \cdot
\bm{\nabla} \bm{\nabla}^2 \, \delta^3 ({\bm{x}}) \right] 
}
{|{\bm{x}} - {\bm{r}}|} \\
 & \quad -  e \frac{\bm{d}_0 \cdot \bm{\nabla}}{Z}
 \int \! d^3x \, \frac{\left[ Z \, \delta^3 (\bm{x}) + \frac{1}{6} Z \langle r^2
\rangle_{\rm ch} \bm{\nabla}^2 \, \delta^3 ({\bm{x}}) \right]}{|{\bm{x}} -
{\bm{r}}|} + \cdots \\
&= 4\pi \bm{S} \cdot \bm{\nabla} \delta^3(\bm{r}) \,
\end{aligned}
\end{equation}
The vector Schiff moment $\bm{S}$ is given by
\begin{equation}
\label{eq:Schiffdef}
\bm{S} \equiv \frac{|e|}{10} \left( \bm{O}_0 - \frac{5}{3} 
\braket{r^2}_{\text{ch}} \bm{d}_0 \right) \,. 
\end{equation}
Thus, the coupling of the symmetry-violating part of the nuclear charge
distribution to the atomic electrons is through the Schiff moment.   The result
(\ref{eq:Schiffdef}) is of order $R_N^3$, because the terms of order $R_N$
cancel in Eq.\ \ref{eq:Del}.

As we've seen, the density $\rho(x)$ comes ultimately from the nuclear
ground-state wave function, and one can easily write an operator whose
expectation value is the Schiff moment (using the exact same symbol,
unfortunately).  If one neglects properties of nucleons beyond their charges,
the vector Schiff operator is
\begin{equation}
\label{eq:Schiff-op}
\bm{S^{\text{ch}}} = \frac{|e|}{10} \sum_{i=1}^{Z}\left( r_i^2 - \frac{5}{3}
\braket{r^2}_\text{ch}  \right) \bm{r}_i \,,
\end{equation}
where the sum is over protons.  To treat the effects of dipole moments for the
nucleons, one must add to the Schiff operator a ``nucleon'' piece
\begin{equation}
\label{eq:Schiffint}
\bm{S}^n = \frac{1}{6} \sum_{i=1}^A \left( r_i^2 -
\braket{r^2}_\text{ch}\right) \bm{d}_i
\end{equation}
where now the sum is over all nucleons and the $\bm{d}_i$ are operators that, by
the Wigner-Eckart theorem, must have the form $\bm{d}_i = D_i \bm{\sigma}_i$.
The coefficients $D_i$ must have one value for all protons $i$ and another
(possibly the same) for all neutrons $i$.   Both the above equations omit a
small piece with quadrupole character and 
a term of order $(Z \alpha)^2$ that is due to relativity in electron wave
functions \cite{Flambaum2002} as well, perhaps, as more subtle electron-nucleus
interactions \cite{liu07}.

Because $\bm{S}^{\text{ch}}$ and $\bm{S}^{n}$ are vectors, a nonzero Schiff
moment requires the nuclear ground state to actually be a multiplet
$\ket{\tilde{g};J,M}$ of $2J+1$ states with different values for the $z$
projection of the total angular momentum $J \geq 1/2$.  (Here $g$ stands for
``ground state'' and the tilde indicates the presence of $V_{PT}$ in the
Hamiltonian.) In such a case, one \textit{defines} the Schiff moment to be the
expectation value of the $z$ component of the total Schiff operator $\bm{S}
\equiv \bm{S}^{\text{ch}} + \bm{S}^n$ in the state with $M=J$, viz.,
\begin{equation}
\label{eq:Schiff-mom-def}
S = \braket{S_z}_{\tilde{g};JJ} \,,
\end{equation}
where the subscripts after the semicolon indicate the values of the total
angular momentum and its $z$ projection.  In perturbation theory in $V_{PT}$,
which is essentially exact given the weakness of that interaction, the Schiff
moment of the nuclear ground-state multiplet $\ket{\tilde{g};J,M}$ that without
the perturbation becomes $\ket{g;J^{\pi},M}$ (the ``g'' has no tilde and the
states have good parity $\pi$) can be written as 
\begin{equation}
\label{eq:Schiff-perturbation}
S = \sum_{n} \frac{\braket{g|S_z|n}_{J,J}
\braket{n|V_{PT}|g}_{J,J}}{E_g - E_n} + c.c \,,  
\end{equation}
where the subscripts on the matrix elements contain the values of quantum
numbers that are well defined in both the bra and ket.  Because $S$ is linear in
$V_{PT}$, one can write, at leading order in $\chi$EFT,
\begin{equation}
\label{eq:Schiff-coeffs}
S = a_0 g \bar{g}_0 + a_1 g \bar{g}_1 + a_2 g \bar{g}_2 + A_1 \bar{C}_1
+ A_2 \bar{C}_2 + a_p d_p + a_n d_n\,, 
\end{equation}
Here $d_p$ and $d_n$ are the nucleon dipole moments and the $a$'s and $A$'s
contain all the nuclear physics.  Given values for the $V_{PT}$ couplings and
the nucleon dipole moments, these coefficients determine the Schiff moment, and
it is the job of nuclear-structure theory to calculate them.  Because of Schiff
screening, experiments on free neutrons are more sensitive to a neutron EDM than
are experiments in atoms and molecules.  As a result, we will focus almost
exclusively on $S^{\rm ch}$, the part of the Schiff moment caused by $V_{PT}$,
and in particular on T-violating pion exchange in $V_{PT}$, i.e., on $a_0, a_1$,
and $a_2$. 

One implication of our derivation of the Schiff theorem above is that an atomic
EDM induced by the nucleus should be a factor of order $\left( \frac{R_N}{R_A}
\right)^2 \approx 10^{-8}--10^{-9}$ smaller than it would be without screening.
In reality, though, the screening is not so strong. The large charge of heavy
nuclei concentrates electrons at the nucleus, an effect that is enhanced by
relativity, with the net result that the atomic EDM is reduced by the much
smaller factor of about $10^{-3}$.  That's still a lot of suppression, but it
can be overcome by experimental ingenuity. 

It's worth noting that at higher order in the multipole expansion, other
electric moments, such as the electric octupole, enter the expression for
$\bar{H}_{\rm atom}$.   (The Schiff moment itself has a small correction
involving the nuclear quadrupole moment.)  Moreover, the nuclear current plays a
role as well as the charge, and when the current's effect on electrons is worked
out, one finds that the magnetic quadrupole moment is important
\cite{khriplovich97,Flambaum94,Flambaum14,Lackenby18}.  Magnetic moments are
unscreened and as a result, in nuclei with $J \geq 1$ and sufficient electron
angular momentum, the atomic EDM induced by the nuclear magnetic quadrupole is
usually larger, by an order of magnitude or more, than that induced by the
Schiff moment.  Magnetic quadrupole moments thus deserve a review of their own,
or at least more space in a review like this one.  Thus far, however, attempts
to compute them have been much fewer, and so, while bearing in mind that in some
experiments their effects will be the most important, they will not be
considered further here. 

\subsection{Experiments in Atoms and Molecules}
\label{sec:exps}

In heavy atoms, electron EDMs are not screened because they are in no way
confined to a small sub-volume and the electronic motion is sufficiently
relativistic.   In paramagnetic atoms, which have unpaired electrons, electron
EDMs can be greatly amplified.  These atoms, however, offer no advantages for
detecting nuclear CP violation unless it is in the form of a symmetry-violating
nucleus-electron interaction, and will not discussed.

Diamagnetic atoms do not accentuate electron EDMs and so are useful for
detecting CP violation within the nucleus.  Some of the best limits on atomic
EDMs come from experiments in diamagnetic atoms.  The most sensitive
\cite{Graner16,Graner17}, on the atom containing the isotope $^{199}$Hg, has
produced the result $|d| < 7.4 \times 10^{-30} \, e\, {\rm cm}$.  A strong limit
also exists in $^{129}$Xe: $|d| < 1.4 \times 10^{-27}\, e\, {\rm cm}$
\cite{Sachdeva19}.  Experiments on the octupole-deformed isotopes $^{225}$Ra and
$^{223}$Ra have been underway for some time \cite{Parker2015,Bishof16} and, for
reasons to be presented, have the potential to be more sensitive than
experiments in isotopes with less exotic shapes.  
   
In recent years, attention has moved to polar molecules, which generate internal
electric fields that can be aligned with an external field to magnify its
effects on EDMs.  Such molecules can be either paramagnetic
\cite{Andreev18,Cairncross17,Hudson:2011zz} or diamagnetic
\cite{Grasdijk21,Malika20}, and thus be most sensitive either to electron EDMs
or nuclear Schiff moments (or magnetic quadrupole moments).  In the future, some
of the latter will leverage the octupole deformation of one of the molecule's
atoms (see Ref.\ \cite{Arrowsmith-Kron24} and references therein).  The field's
rapidly improving ability to cool and control atoms and molecules is exciting.

The problem for theory, of course, is extracting statements about the strength
and likely source of CP violation from the results of any of these experiments.
For nuclear-structure theory, that means the computation of Schiff moments, the
problem to which the rest of this paper is devoted.
 
\section{CALCULATIONS OF SCHIFF MOMENTS}
\label{sec:calcs}

\subsection{Simple Estimate}
\label{sec:est}

The simplest nuclear model in which to estimate the Schiff moment induced by the
interaction $V_{PT}$ is one in which the strong nucleon-nucleon interaction $V$
can be represented by a mean-field $U_0$ and nucleons occupy simple spherical
single-particle levels produced by the mean field.  If one assumes that
excitations of the nuclear core are not important in Eq.\
\eqref{eq:Schiff-perturbation}, then one can also replace the T-violating
potential $V_{PT}$ by a mean field.  The simplest way to do that is to pretend
that pions are heavy compared to an inverse nuclear radius (a gross
approximation, but one that usually doesn't cause large errors in finite model
spaces) so that $Y(r) \rightarrow 1/m_\pi^2 \delta^3(\bm{r})$; the spatial form
of $V_{PT}^\pi$ then becomes the same as that of $V_{PT}^\delta$.  Finally, with
the assumptions that the neutron and proton densities are proportional, that the
spin density is negligible compared to the number density, and that exchange
terms in in $V_{PT}^\pi$ (as well as all terms in $V_{PT}^\delta$) are
unimportant, one arrives at the one-body parity- and time-reversal-violating
potential \cite{Flambaum:1984fb,flambaum86,khriplovich97}: 
\begin{equation}
\label{eq:simple-mf-pot}
U_{PT} = 
\frac{\varepsilon}{M_N m_\pi^2} 
\bm{\sigma} \cdot \bm{\nabla} \rho \tau_z \,,
\end{equation}
where $\rho$ is the total nuclear density
and 
\begin{equation}
\label{eq:etaval}
\varepsilon = \frac{g}{2} 
\left[ \left( \frac{N-Z}{A} \right)
\left( \bar{g}_0 + 2 \bar{g}_2 \right) - \bar{g}_1 \right] \,. 
\end{equation}

In spherical odd-$A$ nuclei, in the crudest approximation, one can assume that
all the nucleons but the last one of the odd system (neutrons or protons) form
an inert ``core'' with total angular momentum zero.  If one makes the final
simplifications that the strong nuclear mean-field $U_0$ is dominated by a
spin-independent part and that $\rho(r)$ and $U_0(r)$ are proportional, one can
(as in some proofs of the Schiff theorem) exploit the fact that the perturbing
Hamiltonian is proportional to $[\bm{\sigma} \cdot \bm{p}, U_0] = -i \bm{\sigma}
\cdot \bm{\nabla} U_0$ to analytically evaluate the sum in the
perturbation-theory expression in Eq.\ \ref{eq:Schiff-perturbation} for the
state of the last (valence) nucleon, obtaining 
\begin{equation}
\label{eq:perturbedwf}
\ket{\tilde{\psi}_{lj}} = \left( 1 + i \varepsilon
\frac{\rho(0)}{ M_N m_\pi^2 U_0(0)} 
\bm{\sigma} \cdot \bm{p} \right) \ket{\psi_{lj}} \,,
\end{equation}   
where $\ket{\psi_{lj}}$ is the valence eigenstate with orbital angular momentum
$l$ and total angular momentum $j$ of the strong mean-field Hamiltonian.  With
this simple state, an estimate for $\rho(0)/U_0(0)$, and the further not totally
unreasonable assumption that the $\braket{\psi|r^2|\psi} \approx \frac{3}{5}
R^2$, with $R = 1.1 \text{fm} \times A^{1/3}$, Ref.\ \cite{Flambaum:1984fb}
finds that the Schiff moment is zero for nuclei with an odd neutron and,
translated into our notation, has a value of 
\begin{equation}
\label{eq:Schiff-valence-simple}
S^{\rm ch} \approx |e| \frac{\left[ 1 \pm \left( j + \frac{1}{2} \right) \right]}{j + 1} 
A^{2/3} \times 10^{-2} \varepsilon \ \text{fm}^{3} \,, 
\end{equation}
for nuclei with an odd proton, with the $\pm$ corresponding to $l = j \pm 1/2$.
One can extend this expression to deformed nuclei and relax some of the
assumptions that enter it, but I will simply present it as a very rough generic
estimate for Schiff moments (though the moments in odd-neutron nuclei are not in
reality systematically much smaller than those in odd-proton nuclei).  One
salient fact is the growth with $A$.
 
In nuclei that are not octupole deformed --- such deformation is a special case
that we'll address later --- it is possible to identify another simple physical
phenomenon that affects Schiff moments: core polarization.  The Schiff operator
is a proton-only version of the isoscalar dipole operator, the experimental
excitation spectrum of which has been studied occasionally \cite{Garg99,
uchida03, clark01}.  Much of the operator's strength --- the squared matrix
element of the operator from the ground state to a given excited state --- lies
in a giant resonance above 20 MeV.  Giant resonances suck strength away from low
energies; in our context, a Schiff resonance above 20 MeV will reduce the
ground-state Schiff moment (for a simple discussion, see Ref.\
\cite{Flambaum1986}.) Quantifying the degree of reduction requires more careful
calculations, calculations which will be discussed later.  For now, it is enough
to note that the simple single-particle result for protons above is probably too
large in most nuclei.  The next section, turns to the natural generalization of
the single-particle picture, the shell model.

\subsection{Phenomenological Shell Model}
\label{sec:shell}

The nuclear shell model \cite{Lawson80,Caurier05} has been applied to light,
medium-mass, and heavy nuclei.  It can work in deformed nuclei but in heavy
deformed nuclei, the computational requirements can become extreme and the model
is less often employed.  For Schiff moments, only the not too deformed isotopes
$^{129}$Xe and $^{199}$Hg have been treated in a version of the model
\cite{Yanase20,Yanase23}. 

The shell model is a generalization of the single-particle picture that has a
number of particles occupying a few valence harmonic-oscillator orbitals rather
than just one particle occupying a single orbital.  To obtain a system's energy
eigenstates, one must diagonalize a Hamiltonian in the space constructed from
the valence orbitals.  But because that space is only a small fraction of the
full many-particle Hilbert space, the Hamiltonian that gives correct energies is
different from the bare nuclear interaction.  Methods for deriving this
``effective interaction'' exist \cite{kuo91,hjorth-jensen95,tsukiyama12,} but in
the phenomenological version of the model, they are implemented only
approximately or not at all, and at least parts of the effective interaction are
fit to data in or near the nuclei under consideration.  Ref.\ \cite{Yanase20}
works with interactions that begin with Brueckner $G$ matrices
\cite{Goldstone57} from phenomenological nucleon-nucleon potentials
\cite{Kaya19,Utsuno14,Herling72,Blomqvist84} and adjust particular matrix
elements to spectra.  In principle, other operators, such as the Schiff
operator, should also be replaced by effective versions.  In the absence of
measured Schiff moments, of course, that last step is difficult.

Even without such difficulties, using the shell model to accurately compute
Schiff moments is a tall order.  The unperturbed shell-model ground state
$\ket{g}$ and low-lying excited states are determined by diagonalizing the
effective Hamiltonian; the model often reproduces energies, electromagnetic
moments, and transition rates exceptionally well.  But to evaluate the
perturbative sum in Eq.\ \ref{eq:Schiff-perturbation} one must include states
$n$ that lie well above and below the valence shell.  The Schiff operator
contains three powers of $r$, and as a consequence, its largest single-particle
matrix elements are between states in the core or valence shell and others that
lie three oscillator shells higher or lower.  A coherent combination of these
states make up the giant isoscalar/Schiff resonance discussed earlier.  The
energies and structures of such states can be obtained only very approximately
in the shell model, which almost by definition focuses on the valence shell.
Refs.\ \cite{Yanase20,Yanase23}, following Ref.\ \cite{Teruya17}, take those
states to be the result of orthogonalizing individual particle-hole excitations
of the ground state.  In other words, the excitations are taken to be
orthogonalized states with the schematic form $a^\dag_i a_j \ket{g}$ and energy
$\varepsilon_i - \varepsilon_j \gtrsim \hbar \omega$, where the $\varepsilon$'s
are singe-particle energies and $\hbar \omega$ is the energy difference between
shells.  The effects of such excitations are found to be small, but the error
introduced by the simplified treatment is hard to quantify.  Ref.\
\cite{Teruya17} makes an attempt to examine the error in the energy denominators
(and claims that it is small) but does not examine the error in the matrix
elements that make up the numerators.  The simple intermediate states that are
assumed to be eigenvectors in this approach almost in reality mix, both with one
another and with more complicated multi-particle multi-hole states that are
completely absent.  The random-phase approximation (RPA) and generalizations
discussed in the next section explicitly capture at least some of this physics,
which is the core polarization mentioned earlier.   The results of shell-model
Schiff-moment calculations appear after a discussion of these other methods,
which is next.

\subsection{Random Phase Approximation and Density-Functional Theory in
Spherical or Symmetrically Deformed Nuclei}
\label{sec:rpa}

Nuclear density functional theory (DFT) is a variation of the Kohn-Sham approach
used in atomic, molecular and condensed-matter theory.  The energy-density
functionals (EDFs) are obtained mainly by writing the contribution to the
mean-field energy of a phenomenological nucleon-nucleon interaction as an
integral involving the density matrix.  If, like Skyrme interactions
\cite{Skyrme58,Schunck19}, the potential has a range of zero, one ends up with a
semi-local function of the density (local in the density and its derivatives
except for the Coulomb piece, which comes from a long-range interaction).  If
the potential has a non-zero range, then only the direct (Hartree) part is local
without further approximation.  Once constructed, the functionals can be
modified in ways that don't correspond to the mean-field expectation value of
any underlying interaction.  In spirit and much of practice, however,
calculations with EDFs are variations on mean-field theory, either the
Hartree-Fock (HF) approximation or its generalization, the
Hartree-Fock-Bogoliubov (HFB) approximation that includes pairing correlations
at the expense of breaking particle-number conservation.

The natural treatment of excited states within mean-field theory is the RPA or,
within HFB, the quasiparticle random-phase approximation (QRPA).  The usual
``matrix'' versions of the approaches involve the diagonalization of the
residual Hamiltonian --- the piece not incorporated into the average field ---
in a space consisting of a one-particle one-hole (or two-quasiparticle for the
QRPA) and one-hole one-particle (two-quasihole) excitations of a not fully
specified ground state.  (It's hard to describe the method in a single sentence;
please see, e.g., Refs.\ \cite{rin04,Schunck19}) The RPA energies and transition
matrix elements turn out to be the same as one gets by computing the linear
response of the Hartree-Fock ground state to a time-dependent perturbation, with
the restriction that time-dependent state always remain a Slater determinant or
quasiparticle vacuum.  Excitation energies in this second picture correspond to
poles in the response function and transition strengths to residues.  For
time-independent quantities such as the Schiff moment, in the case for which the
energy functional is just the mean-field expectation value of a Skyrme
Hamiltonian, the method for computing any of the coefficients $a_i$ is simply to
solve the mean field equations associated with the full Hamiltonian,
\begin{equation}
\label{eq:Hamiltonian}
H = H_{\text{Skyrme}} + \lambda  V_{PT} \,,
\end{equation}
with the corresponding $\bar{g}_{i}$ set to 1 and the others omitted.  The
dimensionless quantity $\lambda$ must be large enough to have a numerical effect
but still small enough so that first-order perturbation theory is accurate.
Once one has the solution, one evaluates $\braket{S_z /\lambda}$ directly as in
Eq.\ \ref{eq:Schiff-mom-def}.  It is not hard to show \cite{ban10} that the
resulting Schiff moment is the same as that produced by the perturbative
expression in Eq.\ \ref{eq:Schiff-perturbation} if the transition matrix
elements in the latter are evaluated in the matrix (Q)RPA.  

Very roughly speaking, there are two kinds of contributions to Eq.\
\ref{eq:Schiff-perturbation} in these schemes.  In the first, $V_{PT}$ affects
only the last (valence) nucleon, like in the simple estimate above, yielding a
non-zero result only in odd-proton nuclei.  In the second, the last nucleon
interacts with the others, which are paired and constitute a kind of core,
leading to particle-hole excitations of the core.  The unperturbed ground state
gets much of its important structure in a similar way through the strong
interaction of the last nucleon with the rest.  These processes, or more
precisely those that go beyond the excitation of uncorrelated one-particle
one-hole states, are the core polarization already mentioned.  The term perhaps
makes most sense in the version of mean-field theory that includes $V_{PT}$
through the direct evaluation of Eq.\ \ref{eq:Schiff-mom-def}; I've noted that
this approach is is equivalent to the RPA.  In it, the wave function of the core
is polarized by the valence nucleon so that the mean field for the core changes
its shape, spin properties, etc., and develops a Schiff moment itself.  

\begin{table}[b]
\tabcolsep7.5pt \caption{Reduction of Schiff-moment coefficients $a_i$ from the
independent-particle approximation, in the calculations of Ref.\
\cite{dmitriev05}.}
\label{tab:corepol}
\begin{center}
\begin{tabular}{l|ccc}
\hline
& $a_0/a_0^{s.p.} $ & $a_1/a_1^{s.p.} $ & $a_2/a_2^{s.p.}$ \\
   \hline
   $^{199}$Hg & 0.004 & 0.61 & 0.05\\
$^{129}$Xe & 0.13 & 0.10 & 0.08\\
$^{211}$Rn & 0.16 & -0.51 & 0.22\\
$^{213}$Rn & 0.10 & 0.18 & 0.07\\
\hline
\end{tabular}
\end{center}
\end{table}
    
The RPA is a flexible method that can be applied without the fully
self-consistent mean field and Skyrme interactions that make up nuclear DFT.
Refs.\ \cite{dmitriev05,dmitriev03} are good examples of that approach.  To
represent the mean field, the papers use Wood-Saxon potentials and their
surface-peaked generalizations (for the spin-orbit potential) rather than
starting with a two-body interaction and carrying out full HF or HFB
calculations.  For the \textit{residual} two-body force, they use an unrelated
zero-range Landau-Migdal interaction.  The core polarization it causes always
reduces the magnitude of the Schiff moment produced by the last proton alone in
odd-$Z$ nuclei (as in Eq.\ \ref{eq:Schiff-valence-simple}), or by the
approximation to Eq.\ \eqref{eq:Schiff-perturbation} (like that in the last
section) in which the intermediate states are all uncorrelated proton
one-particle-one-hole states in odd-$N$ nuclei.  (Both of these will be referred
to as ``independent-particle'' moments.) Table \ref{tab:corepol} shows the
reduction from independent-particle moments effected by the core polarization in
the calculations of Ref.\ \cite{dmitriev05} in several isotopes either used in
experiments or considered for use.  These are all odd-$N$ nuclei.  Here and
everywhere that follows, the independent-particle moments correspond to those in
a mean-field approximation with genuine single-particle wave functions, not the
estimate in Sec.\ \ref{sec:est}, which involved several additional
simplifications, in particular, the neglect even of uncorrelated proton
particle-hole intermediate states.
 
As we've seen, core polarization is almost entirely absent from the large-scale
shell-model calculations.  A small amount is induced by the presence of more
than one nucleon in the valence shell, but that is not enough to cause much
quenching.

Of course, there have also been fully self-consistent Skyrme QRPA calculations,
both in the matrix version of the theory \cite{jesus05} and the linear-response
version \cite{ban10}, in addition to the Wood-Saxon-based computations just
presented.  The scheme in Ref.\ \cite{jesus05} includes some correlations beyond
those in the QRPA, and the calculation in Ref.\ \cite{ban10}, unlike all the
others in mean-field theory, allows nuclear ground states to be deformed.  Both
sets of computations have been carried out with several different Skyrme
functionals, from which there is no compelling reason to choose a favorite.  The
results of these calculations appear just below.

\subsection{Comparison of Results}
\label{sec:compare}

Table \ref{tab:rn} shows the results from several papers for the coefficients
$a_i$ in Eq.\ \ref{eq:Schiff-coeffs}, in the independent-particle approximation,
the phenomenological RPA of Refs.\ \cite{dmitriev05} and \cite{dmitriev03}, and
the still phenomenological but fully self-consistent Skyrme linear-response
(DFT) calculations for the spherical nucleus $^{211}$Rn.  The line for the last
of these contains ranges because the calculations were done with several Skyrme
functionals.   Not very surprisingly, the last two methods agree fairly well
and, also unsurprisingly, produce numbers that are quite quenched from the
independent-particle results, with so much quenching in the isovector ($a_1$)
channel that the sign of the coefficient changes.

\begin{table}
\tabcolsep7.5pt \caption{The coefficients $a_i$, in units of $|e|$ fm$^3$, in
$^{211}$Rn from several nuclear-structure calculations.} 
\label{tab:rn}
\begin{center}
\begin{tabular}{l|ccc}
\hline
Method & $a_0$ & $a_1$ & $a_2$ \\
\hline
Independent particles \cite{dmitriev05} & 0.12 & 0.12& 0.24\\
Phenomenological RPA \cite{dmitriev05} & 0.0019 & -0.061 & 0.053 \\
Skyrme linear response \cite{ban10} & 0.034 $\leftrightarrow$ 0.042 & -0.0004
$\leftrightarrow$ -0.028 & 0.064 $\leftrightarrow$ 0.078 \\
\hline
\end{tabular}
\end{center}
\end{table}

\begin{table}[b]
\tabcolsep7.5pt \caption{The coefficients $a_i$, in units of $|e|$ fm$^3$, for
$^{199}$Hg from a variety of nuclear-structure calculations.} 
\label{tab:hg}
\begin{center}
\begin{tabular}{l|ccc}
\hline
Method & $a_0$ & $a_1$ & $a_2$ \\
\hline
Independent particles \cite{flambaum86} & 0.087 & 0.087 & 0.174 \\
Pair-truncated shell model \cite{Yoshinaga2018} & 0.011 & 0.014 & 0.033 \\
Pair-truncated shell model \cite{Yoshinaga2020} & 0.017 & -0.016 & 0.066 \\
Large-scale shell model \cite{Yanase20} & 0.080 & 0.078 & 0.15 \\
Phenomenological RPA \cite{dmitriev03},\cite{dmitriev05}&0.00004 &0.055 &0.009\\
Skyrme QRPA \cite{jesus05} & 0.002 $\leftrightarrow$ 0.010 & 0.057
$\leftrightarrow$ 0.090 & 0.011 $\leftrightarrow$ 0.025 \\
Skyrme linear response \cite{ban10} & 0.009 $\leftrightarrow$ 0.041 & -0.027
$\leftrightarrow$ +0.005 & 0.009 $\leftrightarrow$ 0.024 \\
\hline
\end{tabular}
\end{center}
\end{table}

Table \ref{tab:hg} compares the results of more calculations for the crucial
isotope $^{199}$Hg. One can expect the numbers here to be less in accord because
the isotope is unpleasantly complicated.  Mean field calculations find it to be
slightly deformed and very soft \cite{Prassa21,Patra94}, implying that a single
mean field with oscillations around it, as in RPA-like treatments, may not be a
very good approximation.  And to the extent that a single mean field can be
used, it may well correspond to a triaxial shape, to which the RPA-like methods
have not been frequently applied. (The phenomenological RPA of Ref.\
\cite{dmitriev05} and the Skyrme matrix QRPA of Ref. \cite{jesus05} treat the
nucleus as spherical, while the Skyrme linear-response approach of Ref.\
\cite{ban10} allows it to be deformed, but without breaking axial symmetry.) The
shell model includes the dynamics of valence nucleons and is best suited for
low-lying states in such nuclei but suffers, as we have seen, from a simplified
treatment of the intermediate states in Eq.\ \ref{eq:Schiff-perturbation}.  

\begin{figure}[b]
\includegraphics[width=.7\textwidth]{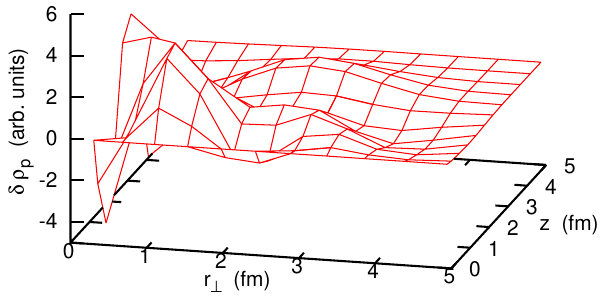}
\caption{ The change in proton density induced by the $\bar{g}_1$ term in
$V_{PT}$, as a function of $r_{\perp} \equiv \sqrt{x^2 + y^2}$ and $z$. The
units are arbitrary because of the arbitrariness in the constant $\bar{g}_1$.
Only a quarter of the density change is shown; it is symmetric in $r_{\perp}$
and antisymmetric in $z$. (Taken from Ref.\ \cite{ban10}.) 
}
\label{fig:hg-density}
\end{figure}

In the table, the large-scale shell model of Ref.\ \cite{Yanase20} produces
coefficients that are almost the same as the independent-particle estimates.
The RPA-like calculations are the most quenched.  For $a_0$ and $a_2$, their
predictions agree fairly well with one another and with those of the
pair-truncated shell model \cite{Yoshinaga2018}, a computationally simpler
version of the shell model that restricts the basis to states involving
collective pairs.  In the isovector channel, however, the linear-response method
is in disagreement with the others, favoring numbers with the opposite sign.
The method includes most of the same physics as the Skyrme QRPA, and the
disagreement is not well understood.  The spread of values is thus indeed large
and one needs more reliable calculations.  That the computation is delicate and
difficult is not surprising.  \textbf{Figure \ref{fig:hg-density}}, from Ref.\
\cite{ban10} shows the computed parity-odd proton density induced by the
$\bar{g}_1$ term in $V_{PT}$, in cylindrical coordinates.  To obtain the Schiff
moment, one must multiply this density change by $ (r^2 - 5/3
\braket{r^2}_{\text{ch}})z$ and integrate.  It's hard to guess even what sign
will result from such a calculation.

Reference \cite{Yanase20} does suggest an explanation for the discrepancy
between the results in different schemes.  It notes that in addition to the
$J^{\pi} = \frac{1}{2}^-$ ground state, $^{199}$Hg has a low-lying excited state
with the same quantum numbers.  The calculated Schiff moment for the excited
state is much smaller than that for the ground state.  The paper's authors
suggest that the mean field found, e.g., in Ref. \cite{ban10} mistakenly
contains a significant admixture of the low-lying excited state.  That
conjecture remains to be explored.

Table \ref{tab:xe} shows results for $^{129}$Xe, an isotope used in some
experiments \cite{Sachdeva19}.  As in $^{199}$Hg, the shell model produces
larger numbers than the other methods, though here even its numbers are quenched
compared to those in the independent-particle picture.  DFT has not been applied
to this isotope.
\begin{table}
\tabcolsep7.5pt \caption{The same as Table \ref{tab:hg}, but for
$^{129}$Xe.} 
\label{tab:xe}
\begin{center}
\begin{tabular}{l|ccc}
\hline
Method & $a_0$ & $a_1$ & $a_2$ \\
\hline
Independent particles \cite{flambaum86} & -0.11 & -0.11 & -0.22 \\
Pair-truncated shell model \cite{Yoshinaga2013} & 0.0005 & -0.004 & 0.0019 \\
Pair-truncated shell model \cite{Teruya17} & 0.0032 & -0.0012 & 0.0042 \\
Large-scale shell model \cite{Yanase20} & -0.038 & -0.041 & -0.081 \\
Phenomenological RPA \cite{dmitriev05}&-0.008 &-0.006 &-0.009\\
\hline
\end{tabular}
\end{center}
\end{table}

\subsection{A Special Case: Octupole Deformed Nuclei}
\label{sec:oct}

\begin{figure}[b]
\includegraphics[width=.3\textwidth]{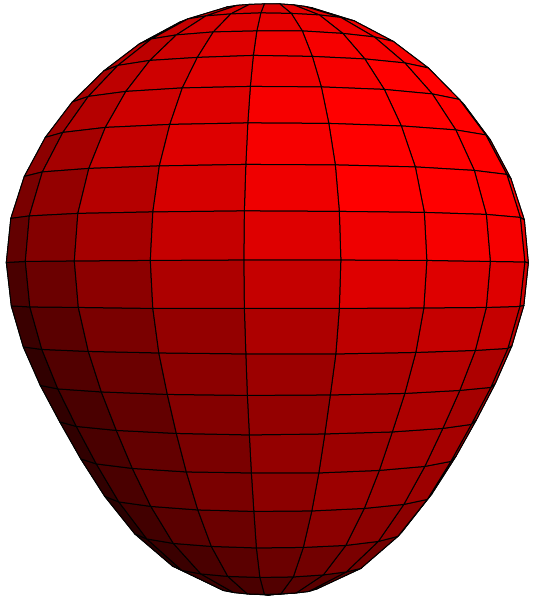}
\caption{Shape of $^{225}$Ra from the calculation of Ref.\ \cite{engel03},
represented by the surface of a uniform body with the same multipole moments as
those of the calculated mass density (taken from Ref.\ \cite{Dobaczewski18}).
}
\label{fig:ra-density}
\end{figure}

There is one class of nuclei --- those with pear-like shapes corresponding to
octupole deformation --- in which Schiff moments are greatly enhanced.  Such
nuclei are rare, but regions of the isotopic chart, particularly the
light-actinide region, contain them.  \textbf{Figure \ref{fig:ra-density}} shows
a Skyrme-HFB calculation \cite{Dobaczewski05} of the surface of the isotope
$^{225}$Ra.  Nearby nuclei are similarly shaped.  

To understand why the asymmetric shape leads to large Schiff moments, note that
each rotational band in the spectrum of a rotating octupole has a partner band
of opposite parity.  In rigid odd-$A$ nuclei, with body-fixed octupole-deformed
mean fields, that statement means that every state has a degenerate partner with
the same intrinsic structure but opposite parity; the two states are projections
onto positive and negative parity of the same intrinsic body-frame state.  Real
nuclei are not perfectly rigid, of course, and in those that are octupole
deformed one member of the doublet is lower in energy than its partner.  The
splitting can be quite small, however, compared to typical nuclear excitation
energies.

The small splitting between partners in a parity doublet means that the partner
of the ground state dominates the sum in Eq.\ \ref{eq:Schiff-perturbation}.  And
the contribution of that state is even larger than the splitting suggests
because the numerator contains $\braket{g|S_z| \bar{g}}_{J,J}$, where the bar
denotes the parity-doublet partner.  This quantity, up to a Clebsch-Gordan
coefficient, is close to the intrinsic matrix element of $\hat{S}_{z}$, which is
the classical Schiff moment of a charge distribution like that in Fig.\
\ref{fig:ra-density}.  The moment is collective because the pear shape of the
nuclear charge density comes from coherent contributions of many single-particle
orbitals.  

In the rigid-rotor limit, for which these statements are exact, and with the
assumption that the contributions of all states but the parity-doublet partner
are negligible, Eq.\ \ref{eq:Schiff-perturbation} becomes 
\begin{equation}
\begin{aligned}
\label{eq:Schiff-intr}
S  & \approx \frac{\braket{g|S_z|\bar{g}}_{J,J}
\braket{\bar{g}|V_{PT}|g}_{J,J}}{E_g - E_{\bar{g}}}\, +\, c.c \\
&  = -2 \frac{J}{J + 1} \frac{\braket{S_z}_\text{int}
\braket{V_{PT}}_\text{int}}{\Delta E} \,, 
\end{aligned}
\end{equation}
where $\Delta E = E_{\bar{g}} -E_g$ and the subscript ``int'' refers to the
intrinsic symmetry-breaking state.  All that must be computed, then, are the
intrinsic ground-state matrix elements of $S_z$ and $V_{PT}$.  Despite the
simplicity of this description, however, an equally simple estimate of the size
of $S$ is not so easy to obtain.  Although $\Delta E$ can be taken from
experiment and $\braket{S_z}_\text{int}$ related to other experiments in a way
to be described shortly, $\braket{V_{PT}}$ is a more delicate quantity.  Results
of more than one kind of calculation nonetheless show that the Schiff moment of
$^{225}$Ra is much larger than that of $^{199}$Hg. 

Table \ref{tab:ra} shows the results of these calculations for $^{225}$Ra. The
first is from a particle-rotor model in Ref.\ \cite{spevak97}, the authors of
which first pointed out the octupole enhancement \cite{spevak95,Auerbach96}.
They used an octupole-deformed Wood-Saxon potential from the
semi-self-consistent work in Ref.\ \cite{Cwiok91}, to calculate the intrinsic
matrix elements in the last line of Eq.  \ref{eq:Schiff-intr} and took $\Delta E
= 55$ keV from experiment.  The second set of results is from Skyrme HFB
calculations of the same quantities, with $\Delta E$ again taken from
experiment.  The range in the table corresponds, as for $^{199}$Hg, to the
spread in the predictions of several Skyrme functionals. 

\begin{table}[t]
\tabcolsep7.5pt \caption{The coefficients $a_i$, in units of $|e|$ fm$^3$ for
$^{225}$Ra from a several nuclear-structure calculations.} 
\label{tab:ra}
\begin{center}
\begin{tabular}{l|ccc}
\hline
Method & $a_0$ & $a_1$ & $a_2$ \\
\hline
Phen.\ octupole deformation \cite{spevak97} & 2.7 & -13.5 & 5.4 \\
Skyrme DFT \cite{Dobaczewski05} & 1.0 $\leftrightarrow$ 4.7& -6.0
$\leftrightarrow$-21.5 & 3.9 $\leftrightarrow$ 11.0 \\
Constrained Skyrme DFT \cite{Dobaczewski18} & -0.4 $\leftrightarrow$ 0.8& -2.0
$\leftrightarrow$-8.0 & 1.8 $\leftrightarrow$ 4.8 \\
\hline
\end{tabular}
\end{center}
\end{table}

\begin{figure}[t]
\includegraphics[width=.6\textwidth]{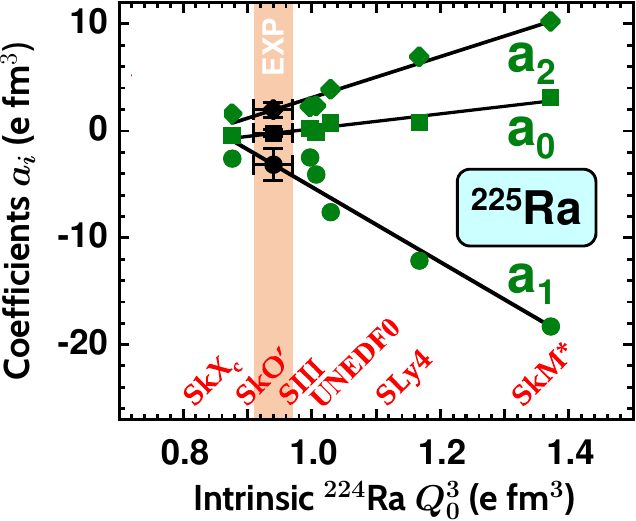}
\caption{Coefficients $a_i$, in units of $|e|$ fm$^3$, in $^{225}$Ra for six
Skyrme functionals and propagated to the measured octupole moment in $^{224}$Ra}
\label{fig:ra-constrain}
\end{figure}

The source of the last line, from Ref.\ \cite{Dobaczewski18}, requires a
slightly longer description.  That paper shows that the intrinsic Schiff moment
in Eq.\ \ref{eq:Schiff-intr} is very tightly correlated with the intrinsic
octupole moment associated with the operator $\sum_{i=1}^{Z} r_i^3
Y^3_0(\theta_i,\varphi_i)$; the predictions for the two quantities by all Skyrme
functionals lie on a straight line when plotted versus one another.  The
intrinsic expectation value of $V_{PT}$, the other quantity in Eq.\
\ref{eq:Schiff-intr} is not very correlated with the octupole moment, but the
product of the two factors retains a correlation.  \textbf{Figure
\ref{fig:ra-constrain}} plots the coefficients $a_i$ produced by HFB
calculations with six Skyrme functionals, along with the experimental intrinsic
octupole moment, obtained from measured E2 and E3 transition strengths
\cite{gaffney13}.  Although the correlation is not perfect, it exists, and by
interpolating the results so that they reproduce the measured octupole moment,
one obtains predictions for the coefficients.  The last line in Tab.\
\ref{tab:ra} comes from a similar analysis in which the measured octupole
moments in both $^{224}$Ra and $^{226}$Ra \cite{Wollersheim93} are used in the
fit.  

A comparison of Tab.\ \ref{tab:ra} with Tab.\ \ref{tab:hg} indeed reveals a
large enhancement for $^{225}$Ra .  Other octupole-deformed light actinides also
have enhanced Schiff moments, and polar molecules containing them have recently
become attractive for experiment.

The first and second lines of Tab.\ \ref{tab:ra} are different from what was
reported in the review article, Ref.\ \cite{Engel13}.  The reasons are a
mistaken expression for $\varepsilon$ in Eq.\ \ref{eq:etaval} in that paper, and
sign errors in Ref.\ \cite{Dobaczewski05}.
%


\section{THE NEAR FUTURE: AB INITIO CALCULATIONS}
\label{sec:future}

We have already seen that $\chi$EFT determines the form of $V_{PT}$.  It will
also allow us to identify corrections to the Schiff operators in Eqs.\
\ref{eq:Schiff-op} and \ref{eq:Schiffint}.  Such corrections will be much
smaller, however, than the inaccuracies in calculated Schiff moments coming from
approximate solutions of the kind presented above to the nuclear many-body
problem.  Here we examine ways in which those solutions can become more accurate
in the next few years.  The program laid out parallels an effort, already
underway, to improve the computation of nuclear matrix elements for neutrinoless
double-beta decay \cite{Cirigliano2022}. 

\subsection{Ab Initio Many-Body Methods}
\label{sec:abinit}

The development of $\chi$EFT and increases in the power of computers over the
last 20 years have moved \textit{ab initio} computation to the forefront of
nuclear-structure theory.  The term \textit{ab initio} refers to calculations
that start from Hamiltonians and other operators with forms specified by
$\chi$EFT (usually), and coefficients most often fit to data from
nucleon-nucleon scattering and properties of two- and three-nucleon systems such
as the deuteron, triton, and $^{3}$He.  Once one has a strong and $PT$-violating
Hamiltonian, the task of interest is to diagonalize the sum of the two of the
two and compute the expectation value of $S_z$, either through Eq.\
\ref{eq:Schiff-mom-def} or Eq.\ \ref{eq:Schiff-perturbation}. 
 
Quantum Monte Carlo \cite{Nightingale98} is a scheme for evaluating the
many-dimensional integrals that define a system's energy or time evolution.
Several versions of the method have been applied to nuclear structure
\cite{Carlson15,Gandolfi20}.  Green's function Monte Carlo \cite{Carlson87} is
the most accurate but is limited to nuclei lighter than about $A \approx 16$;
the notorious sign problem leads to an exponential scaling of computing time
with system size and limits the ability of many kinds of Monte Carlo to work in
heavy nuclei.  There, other more approximate methods with polynomial scaling
have proved useful.  These methods, while not exact, are ``systematically
improvable,'' meaning that they truncate correlations and can be made
arbitrarily accurate, if more costly, by relaxing the truncation.  Two such
methods promise better calculations of Schiff moments: the In-Medium Similarity
Renormalization Group (IMSRG) \cite{Hergert16,Hergert16a}, and the
coupled-cluster method \cite{Shavitt09,dean03}.  Both are discussed below.

\subsection{In-Medium Similarity Renormalization Group}
\label{sec:imsrg}

The IMSRG is a scheme for finding a unitary transformation of the \textit{ab
initio} nuclear Hamiltonian that decouples a low lying state or space of states
from the rest of the many-body Hilbert space.  The idea is to obtain the
transformation in small steps, in a way that is similar to gradient descent.
One does so by solving a differential ``flow'' equation,
\begin{equation}
\label{eq:flow}
\frac{d}{ds} H(s) = \left[ \eta(s),H(s) \right] \,, 
\end{equation}
where $s$ is a time-like flow parameter.  Here $H(0)$ is the Hamiltonian that we
are given. A number of options are available for the ``generator'' $\eta(s)$;
one simple choice comes from dividing the Hamiltonian into a piece $H_d$ that
does not couple the space of low-lying states one cares about to the rest ($d$
stands for ``diagonal''), and a second piece, $H_{od}$, that does and that one
wants to drive to zero.  It's straightforward to show that the generator
\begin{equation}
\label{eq:generator}
\eta(s) = [H_d(s),H_{od}(s)] \,, 
\end{equation}
does the trick, with the effective Hamiltonian in the low-lying space given by
the projection onto that space of $H_d(\infty)$.  Other generators that do the
same job are better in practice because the flow equations that they produce are
not as stiff \cite{Hergert16}.

No matter what generator one chooses, however, Eq. \ref{eq:flow} is too hard to
solve exactly. $H(0)$ and $\eta(0)$ both contain two-body pieces and sometimes
three-body pieces, and as a result $H(s)$ develops up to $A$-body pieces.  One
must truncate at the two-body or three-body level, after normal-ordering with
respect to the ``reference'' state or ensemble of states that one wishes to
decouple.  (The normal ordering is crucial; it incorporates the parts of
higher-body interactions that have a non-zero reference expectation value.).
The simplest version of the method decouples a single reference Slater
determinant.  The valence space version (denoted by the acronym VS-IMSRG)
decouples all the states constructed from valence nucleons in one or two shells
\cite{Stroberg17}.  Finally, the ``In-Medium Generator-Coordinate Method''
(IM-GCM) approximately decouples a single correlated state or ensemble
constructed through the use of generator coordinates such as deformation
parameters and pairing gaps \cite{Yao18}.  The VS-IMSRG promises accurate Schiff
moments in soft nearly spherical nuclei such as $^{199}$Hg and the IM-GCM does
the same in octupole-deformed nuclei such as $^{225}$Ra.  The two approaches
will be considered in turn.  

\subsubsection{Valence-Space IMSRG and Near-Spherical Nuclei}
\label{sec:vs-imsrg}
We have seen that the shell model can describe the low-lying structure of
$^{129}$Xe and $^{199}$Hg, but has trouble including the core polarization
induced by high-energy intermediate states in Eq.\ \ref{eq:Schiff-perturbation}.
The IMSRG, however, can incorporate physics at high energies into valence-space
Hamiltonians and moments.  If, as in Eq.\ \eqref{eq:Hamiltonian}, one writes the
full Hamiltonian in the form
\begin{equation}
H(s) = H_0(s) + \lambda V_{PT}(s)  
\end{equation}
where $H_0$ is the strong Hamiltonian, and write the generator as 
\begin{equation}
\eta(s) = \eta_0(s) + \lambda \eta_{PT}(s) \,, 
\end{equation}
one can expand the flow equations to first order in $\lambda$ to obtain 
\begin{equation}
\label{eq:H0FLow}
\frac{dH_0(s)}{ds} = [\eta_0(s) ,H_0(s)] 
\end{equation}
and
\begin{equation} \label{eq:VtildeFlow}
\frac{dV_{PT}(s)}{ds} = [\eta_{PT}(s),H_0(s)] + [\eta_0(s),V_{PT}(s)].
\end{equation}
The first of these equations is just the usual IMSRG flow; the second evolves
$V_{PT}$ to an effective version for use in the valence shell.  After solving
these, one can transform the Schiff operator in a similar way and use the
effective valence Hamiltonian and Schiff operator to compute the Schiff moment.
The effects of the intermediate states outside the model space should be
captured by these effective operators. 

When the commutators in the flow equation are truncated at the one-body level,
after normal ordering, the IMSRG with a Slater determinant reference state is
similar to the HF approximation \cite{Hergert16}; it decouples all one-particle
one-hole states from the reference.  Recall from Section \ref{sec:rpa} that the
linear-response DFT computation of the static Schiff moment reduces to a
mean-field calculation with $V_{PT}$ included.  The implication is that the
IMSRG should include all the physics of that calculation, plus additional
correlations from truncating at the two-or-three-body normal-ordered level
instead of the one-body level and using a strong Hamiltonian from $\chi$EFT.  We
anticipate useful \textit{ab initio} Schiff moments in $^{199}$Hg and $^{129}$Xe
from this scheme.  

\subsubsection{IM-GCM and Octupole-Deformed Nuclei}

Pear-shaped nuclei are so exotically deformed that one or two valence shells are
not sufficient even to treat them approximately.  But, as we saw in Section
\ref{sec:oct}, we need only the ground-state and its opposite parity partner to
compute the Schiff moment.  The IM-GCM, which focuses on just a few low-lying
collective states, is thus the version of the IMSRG that is best suited for the
computation.

The IM-GCM applies methods such as symmetry-breaking mean-field theory and the
projection of symmetry-broken states onto their unbroken counterparts that were
developed in nuclear DFT.  (It also allows the mixing of different mean fields,
which is the essence of the generator-coordinate method.)  Thus, just as the
ideas in DFT response calculations can be incorporated into the VS-IMSRG, the
spontaneous parity-breaking DFT calculations described in Sec.\ \ref{sec:oct}
can be generalized in the IM-GCM.  A good IM-GCM calculation will start with a
parity- and rotational-symmetry-breaking HF or HFB calculation, with projection
of the resulting quasiparticle vacuum onto states with the quantum numbers of
the ground state and its partner creating a two-state ensemble that can serve as
the reference for normal ordering. (Ensemble normal ordering is used both in the
VS-IMSRG\cite{Stroberg17} and in the application of the IM-GCM to double-beta
decay \cite{Yao18,Yao20}.)  After creating the reference, one can solve the
corresponding flow equations and use the first line of Eq.\ \ref{eq:Schiff-intr}
to compute the Schiff moment.  Because normal ordering with respect to a
complicated state or ensemble is different from ordering with respect to a
Slater determinant, the flow equations do not fully decouple the reference and
are followed by a re-diagonalization of the Hamiltonian, in a space consisting
of the reference states plus the most important non-reference states; at the
end, one will use Eq.\ \eqref{eq:Schiff-intr} again.  

There are no major obstacles in the path to a computation of the Schiff moment
of $^{225}$Ra.

\subsection{Coupled-Cluster Method}
\label{sec:cc}

The coupled-cluster method has a long history, beginning in nuclear physics
\cite{Kummel78}, then undergoing extensive development in atomic and molecular
physics/chemistry \cite{Crawford07}, and finally returning to nuclear physics
\cite{dean03,hagen14} some 20 years ago.   Within nuclear physics, it has been
applied even in nuclei as heavy as $^{208}$Pb \cite{Hu22}, and to processes
ranging from photo-absorption \cite{Bacca13} to double-beta decay
\cite{Novario2021}.   

The basic idea of the approach is to write the nuclear ground state in the
completely general form,
\begin{equation}
\label{eq:cc}
\ket{g} = e^{T} \ket{\Phi} \,,
\end{equation}
where $\ket{\Phi}$ is a Slater determinant and
\begin{equation}
\label{eq:Tcc}
T = \sum_{mi} t^m_i a^\dag_m a_i + \sum_{mnij} t^{mn}_{ij} a^\dag_m a^\dag_n
a_i a_j + \dots \,,
\end{equation}
where $m,n, \dots$ label particle orbits and $i,j, \dots$ hole orbits and the
$t$'s are amplitudes. The expansion for $T$ is truncated at the level indicated
above or, sometimes, with three-body operators included as well.  The
exponentiation of $T$ in Eq.\ \ref{eq:cc}, even when truncated, creates
comprehensive correlations similar to those induced by the IMSRG flow equations.

Initially, nuclear coupled-cluster theory was used only in spherical isotopes,
but recently, practitioners have developed a deformed-basis version of the
method, with the operators in Eq.\ \ref{eq:Tcc} creating and destroying deformed
orbitals, and the ability to project rotational-symmetry-breaking states onto
those with good angular momentum \cite{Hagen22}.  Breaking and restoring parity
symmetry will be a small step on top of that and will allow a computation of
matrix elements of operators between members of a parity doublet.  To mirror
what is possible in the IM-GCM, it might also be necessary to break the U(1)
symmetry that corresponds to particle-number conservation, as in HFB, in the
theory's creation and annihilation operators.  A quasiparticle coupled-clusters
theory has been developed \cite{Henderson14,Signoracci15}, but not yet combined
with the version that allows deformation.  As with the IM-GCM, however, there
appear to be no insurmountable obstacles to an accurate calculation of the
Schiff moment of $^{225}$Ra. 

\section{Conclusion}

This article has been meant to show both the importance of computing nuclear
Schiff moments and the difficulty in doing so well.  A variety of traditional
methods have been applied to the problem, with success that is hard to quantify.
\textit{Ab initio} methods promise more accurate results.  Though the problem of
estimating uncertainty hasn't been addressed here, Ref.\ \cite{Cirigliano2022}
takes up the issue for calculations of neutrinoless double-beta decay, and a
program similar to the one outlined there would be useful for Schiff moments as
well.  All the tools exist to advance the computation of these vital nuclear
moments substantially in the next few years.

\section*{DISCLOSURE STATEMENT}
The author is not aware of any affiliations, memberships, funding, or
financial holdings that might be perceived as affecting the objectivity of this
review. 

\section*{ACKNOWLEDGMENTS}
The author acknowledges the support of the US Department of Energy, Office of
Nuclear Physics, under Grant No.\ DE-FG02-97ER41019, and coauthors S.\ Ban, P.\
Becker, M.\ Bender, J.\ Dobaczewski, J.\ Friar, A.C.\ Hayes, J.H.\ de Jesus, M.\
Kortelainen, C.P. Liu, M.J.\ Ramsey-Musolf, P.\ Olbratowski, and U.\ van Kolck,
whose work contributed figures and results to this article.  He also thanks
collaborators A.\ Belley, D.\ Kekejian, B.\ Romeo Zaragozano, and R.\ Stroberg
for enlightening conversations and useful work.  

%


\end{document}